\documentclass[conference]{IEEEtran}
\IEEEoverridecommandlockouts
\usepackage{amsmath,amsfonts}
\usepackage{algorithmic}
\usepackage{algorithm}
\usepackage{array}
\usepackage[caption=false,font=normalsize,labelfont=sf,textfont=sf]{subfig}
\usepackage{textcomp}
\usepackage{stfloats}
\usepackage{url}
\usepackage{verbatim}
\usepackage{graphicx}
\usepackage{caption}
\usepackage{romannum}
\usepackage{cite}
\usepackage{amsmath}
\usepackage{makecell}
\usepackage{multirow}
\usepackage{xcolor
}
\usepackage{adjustbox}

\begin{document}

\title{Taylor-Expansion-Based Robust Power Flow in Unbalanced Distribution Systems: A Hybrid Data-Aided Method}
\vspace{-5pt}
\author{\IEEEauthorblockN{Sungjoo Chung, Ying Zhang}
\IEEEauthorblockA{\textit{Department of ECE} \\
\textit{Oklahoma State University},
Stillwater, U.S. \\
\{sungjoo.chung;y.zhang@okstate.edu\}}
\and

\IEEEauthorblockN{Zhaoyu Wang}
\IEEEauthorblockA{
\textit{Department of ECE} \\
\textit{Iowa State University},
Ames, U.S. \\
wzy@iastate.edu}

\and
\IEEEauthorblockN{Fei Ding}
\IEEEauthorblockA{
\textit{National Renewable Energy Laboratory}\\
Golden, U.S. \\
fei.ding@nrel.gov}

}




\maketitle
\vspace{-10pt}
\begin{abstract}
Traditional power flow methods often adopt certain assumptions designed for passive balanced distribution systems, thus lacking practicality for unbalanced operation. Moreover, their computation accuracy and efficiency are heavily subject to unknown errors and bad data in measurements or prediction data of distributed energy resources (DERs). To address these issues, this paper proposes a hybrid data-aided robust power flow algorithm in unbalanced distribution systems, which combines Taylor series expansion knowledge with a data-driven regression technique. The proposed method initiates a linearization power flow model to derive an explicitly analytical solution by modified Taylor expansion. To mitigate the approximation loss that surges due to the DER integration and bad data, we further develop a data-aided robust support vector regression approach to estimate the errors efficiently. Comparative analysis in the 13-bus and 123-bus IEEE unbalanced feeders shows that the proposed hybrid algorithm achieves superior computational efficiency, with guaranteed accuracy and robustness against outliers.
\end{abstract}


\section{Introduction}

\IEEEPARstart{P}{ower} flow (PF) is of fundamental importance to secure operation, control, and management of electric distribution systems. Particular characteristics of distribution systems, such as high \(R\)/\(X\) ratios, unbalanced operation, and radial network structures, necessitate the development of dedicated solutions for the PF problem \cite{CIM867133,YZ8753417}. Furthermore, the deepened penetration of distributed energy resources (DERs) underscores the importance of this research for active distribution networks (ADNs) to dispatch these flexible sources. As the PF model is nonlinear and non-convex, notable efforts have been devoted to the design of various numerical methods, e.g., backward-forward sweep (BFS) \cite{kersting2018distribution}.

To facilitate the integration across various optimization tasks and accelerate computational speed, a host of linearization approximation methods \cite{balancedLinearD7031975,Yang7782382,Heidari9406380,BernsteinLinear8332975}  based on certain assumptions have been developed. For instance, a linearized PF method in balanced distribution systems is proposed in \cite{balancedLinearD7031975}, assuming that shunt admittance is negligible. In \cite{Yang7782382}, the assumption of flat voltage profiles, i.e., around 1.0 p.u., are adopted. However, the above assumptions lack congruence with practical conditions, especially in secondary distribution feeders with single- and two-phase line configurations and DER installation. Therefore, sequential efforts have been dedicated to the approximate representation of the PF equations in unbalanced systems  \cite{BernsteinLinear8332975,Heidari9406380}. In \cite{BernsteinLinear8332975}, two algorithms, namely first-order Taylor expansion and fixed-point iteration, are proposed for the linearization. 

Besides, data-driven PF approaches have gained enormous momentum due to the widespread deployment of advanced metering infrastructure (AMI) systems in modern distribution networks, leading to a surge in available historical data. Various machine learning techniques, e.g., regression, are adopted in this domain \cite{Weng9524510, Chen9347375, KplaneRegression9662079,Zhengshuo9626603}. For example, a support vector regression (SVR)-based approach is proposed in \cite{Chen9347375} to address the adverse impacts of outliers on PF calculation. 
Recently, a hybrid learning-aided approach was developed in \cite{Zhengshuo9626603}, and based on PF formulation, typical load conditions and resultant voltages from historical data are utilized to put forth a linear PF method by least squares regression. Nevertheless, the scenarios of high penetration of DER generation have not been investigated. 

Besides, the prevalence of outliers in modern distribution systems poses another challenge to obtaining reliable PF solutions. Anomalous data, recorded by meters, might arise for a variety of reasons, due to meter malfunction, false data injection attacks, or communication interruption \cite{BadData9638014,YZ9144530}. Furthermore, traditional linear regression-based approaches are vulnerable to outliers. Therefore, it is still an open issue how to handle significant uncertainty and unexpected outliers, either of which distinctly impair the validity and efficiency of the existing PF methods. 

To address this issue, this paper proposes a data-aided robust PF algorithm in unbalanced ADNs by combining model-based Taylor series expansion and data-driven SVR techniques. 
The proposed method initiates a linearization PF model to derive an explicitly analytical solution, which is built on a modified Taylor expansion for unbalanced distribution systems. Furthermore, as the linearization errors might surge with dramatic load and DER changes, an auxiliary learning model using a limited number of historical datasets is sequentially constructed for overall accuracy refinement. An SVR-based approach is developed to estimate the approximation errors of the Taylor-based solution and then enhance the solution accuracy through the offset significantly. 
Consequently, the proposed hybrid algorithm exhibits superior computational efficiency and accuracy over several sole model-based and data-driven methods. 
Such desirable performance of the proposed method is independent of the widely used assumptions for traditional passive power systems, e.g., the voltage magnitudes are close to 1.0 p.u. Moreover, case studies demonstrate that the proposed method is robust against possible data outliers and realizes at least 20-fold accuracy enhancement. 

\section{Problem Formulation}

Denote a radial distribution system by a graph $(\mathcal{N},\mathcal{E})$, where $\mathcal{N}:=\{0,1,2,...,n\}$ is the node set and $\mathcal{E} \subset \mathcal{N} \times \mathcal{N}: =\{(i,j)|i,j\in \mathcal{N} \}$ is the edge set. The slack node (substation) is indexed by 0, and the set of other nodes is denoted as $\mathcal{N}\backslash \{0\}$. Given a multi-phase unbalanced distribution network, let $\Phi_i \in \{a,b,c\}$ denote a set of all the existing phases on bus $i$. Then, the total number of the phases on all non-slack nodes of this system is denoted as $\mathcal{M}=\sum_{i \in \mathcal{N}\backslash \{0\}} |\Phi_i|$.

Without loss of the generality, the power injection at bus $k$ through single-, two-, or three-phase connection holds below 
\begin{equation}\label{Power}
\mathbf s_{k}=\mathbf p_k+\mathrm j\mathbf q_k=\mathbf v_k \sum_{m \in N(k)}[\mathbf Y_{mk} (\mathbf v_m-\mathbf v_k)]^*    
\end{equation}
where $\mathbf s_k\in \mathbb C^{|\Phi_i|}$  denotes the complex power at bus $k$, and $N(k)$ is the set of all buses connected to bus $k$; $\mathbf Y_{mk}\in \mathbb C^{|\Phi_i|\times|\Phi_i|}$ denotes the line admittance between nodes $m$ and $k$, and it is calculated from known line impedance and shunt capacitance. Some numerical PF methods such as BFS \cite{kersting2018distribution} are developed to iteratively solve \eqref{Power}.


For the unbalanced distribution network, in the compact form, the multi-phase Ohm's law for all the nodes is given by \cite{ BernsteinLinear8332975}:
\begin{equation}\label{11} 
\begin{bmatrix} \mathbf {I}_0  \\ \mathbf{I}_N \end{bmatrix} =\begin{bmatrix} \mathbf Y_{00} & \mathbf Y_{0N} \\
\mathbf Y_{N0} &  \mathbf Y_{NN} \end{bmatrix}\begin{bmatrix} \mathbf {V}_0  \\ \mathbf{V}_N \end{bmatrix}=\mathbf Y\begin{bmatrix} \mathbf{V}_0  \\ \mathbf{V}_N \end{bmatrix}
\end{equation}
where the subscripts 0 and N denote the slack and non-slack nodes, respectively; the voltage vectors for all non-slack nodes are denoted as $\mathbf{V}_N=\{\mathbf{v}_k\}_{k\in \mathcal N\backslash \{0\}} \in {\mathbb{C}}^{\mathcal{M}}$ and $\mathbf {V}_0 \in {\mathbb{C}}^{3}$;  $\mathbf{Y}$ denotes the nodal admittance matrix for this multi-phase network,  $\mathbf{Y}_{N0} \in {\mathbb{C}}^{\mathcal{M}\times3}$ and $\mathbf{Y}_{NN} \in {\mathbb{C}}^{\mathcal{M}\times \mathcal{M}}$. 


The objective of the PF problem is to solve $\mathbf {V}_N $ based on the following nonlinear power balance equation in the compact form \cite{CIM867133}:
\begin{equation}\label{SI}
\mathbf {S}  =  \mathbf {V}_N \odot \mathbf{I}_N^* = \mathbf {V}_N \odot(\mathbf{Y}_{N0} \mathbf {V}_0+\mathbf{Y}_{NN} \mathbf{V}_N)^*
\end{equation}
where $\mathbf S \in {\mathbb{C}}^{\mathcal{M}}$ denotes the complex power vector, and $(\cdot)^*$ and $\odot$ denote the complex conjugate and element-wise multiplication, respectively. 


\section{Proposed Hybrid SVR-Aided Algorithm}

This section proposes a hybrid Taylor-expansion-based SVR-aided PF algorithm, which marries the physical formulation from the Taylor expansion for unbalanced distribution systems to enhance the computation accuracy and robustness against bad data. The input includes power injection measurements from smart meters and forecasting data of DER generation if there is no meter installed. Built on Taylor series expansion and SVR, the proposed hybrid algorithm is able to simultaneously handle unknown errors and bad data existing in the measurement dataset. 


\subsection{Rotated Complex Taylor Series Expansion}

 \begin{figure}[!t]
   \centering
   \includegraphics[width=0.19\textwidth]{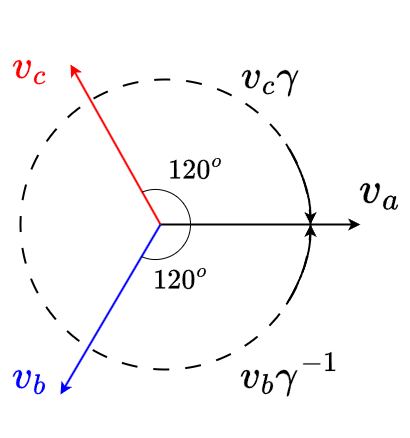}
 \caption{Three-phase rotation for linearization approximation}
 \vspace{-5pt}
 \end{figure}
 
A modified Taylor series-based linear approximation of the PF model is proposed in a multi-phase unbalanced distribution system. First, the nonlinear relationship between $\mathbf V_N$ and $\mathbf S$ in \eqref{SI} is rewritten as:  
\begin{equation}\label{S_U2}
[\mathbf {S} \oslash \mathbf {V}_N]^* = \mathbf{Y}_{N0} \mathbf {V}_0+\mathbf{Y}_{NN} \mathbf{V}_N
\end{equation}
where $\oslash $ denotes the element-wise division. 

Then, a linear approximation for voltage phasors is developed in the per-unit system. 
Define $\Delta v_k=1-v_k$, where $v_k$ is the voltage phasor at bus $k$, and apply the Taylor expansion of the complex-valued $\Delta v_k $ within a narrow range, i.e., $ | \Delta v_k| <1$:
\begin{equation}\label{Taylor}
\frac{1}{v_k}=\frac{1}{1-\Delta v_k}=\sum_{n=0}^{\infty}(\Delta v_k)^n \; \; \; \; \; \;   \forall | \Delta v_k| <1
\end{equation}

The precondition $|\Delta v_{k}| < 1$ allows the proposed algorithm to be applied to a wider range of distribution system operations with deepening DER penetration. This is in contrast with the conventional assumption made for passive distribution networks, which assumes a narrow operating range of voltage magnitudes constrained to float in [0.95,1.05]\cite{Zhengshuo9626603}.

Ignoring the high-order terms ($n \geq 2)$ and substituting  $\Delta v_k$, the following approximation is obtained:
\begin{equation}\label{app}
\frac{1}{v_k} \approx1+\Delta v_k=2-v_k 
\end{equation}

The Taylor expansion-based approximation \eqref{app} is then extended to the three-phase system. Define a phasor rotation operator $\gamma $, 
$\gamma \equiv e^{j\frac{2}{3} \pi } $. 
Rotate the original B-phase voltage counterclockwise by 120 degrees to satisfy $|1-v_k'|<1$, where $v_k'=\gamma^{-1}v_k^b$. Substituting $v_k'$ into \eqref{app}, the following approximation for the original B-phase voltage phasor is derived:
 \begin{equation}\label{PR_Taylor2}
 \frac {1}{v_k'}=(2-v_k')
 \end{equation}
 \begin{equation}
     \frac {1}{v_k^b}=\gamma^{-1} (2-\gamma^{-1} v_k^b)
 \end{equation}


The same phasor rotation technique can be applied to C-phase voltages. Finally, a generalized approximation for multi-phase voltage is derived on bus $k$:
\begin{equation}\label{App3}
    \mathbf 1\oslash \mathbf v_k= \mathbf t_k \odot (\mathbf 2-\mathbf t_k \odot \mathbf v_k) 
\end{equation}
where $\mathbf t_k \in \mathbb{C}^3$, $\mathbf t_k =\{1,\gamma^{-1}, \gamma\}$.

Note that the subset of $\mathbf t_k$ can implement the rotation operation for a two- or single-phase voltage $\mathbf v_k \in \mathbb C^{|\phi_k|}$, which is highly advantageous for multi-phase ADNs.
Then, a Taylor expansion-based linear PF model is obtained by substituting \eqref{App3} into \eqref{S_U2}:
\begin{equation}\label{S_U3}
\mathbf {S}^* \odot\mathcal T \odot (2- \mathcal T \odot \mathbf {V}_N^*) = \mathbf{Y}_{N0} \mathbf {V}_0+\mathbf{Y}_{NN} \mathbf{V}_N
\end{equation}
where $\mathcal{T} \in {\mathbb C^\mathcal{M}}$ denotes the rotation vector for all the buses, and $\mathcal{T}=\{\mathbf t_k\}_{k \in \mathcal N}$.

Rewrite the linearized PF equation \eqref{S_U3} as:
\begin{equation} \label{linearPF}
\mathbf A\mathbf V_N^*+\mathbf{B}\mathbf{V}_N=\mathbf d          
\end{equation}
where $ \mathbf A= \text{diag}(\mathbf S^* \odot \mathcal{T}^2)$, $\mathbf B=\mathbf{Y}_{NN}$, and $\mathbf d= 2\mathbf S^* \odot \mathcal{T}-\mathbf Y_{N0}\mathbf V_0$. 

\subsection{Hybrid SVR-Aided Power Flow Algorithm}
Albeit \eqref{linearPF} offers an explicit linear PF formulation, the approximation errors heavily hinge on different operating conditions that result in diversified voltages on all the buses. To address this issue and enhance the accuracy and robustness against bad data, we further propose a hybrid robust SVR-aided PF method. By estimating the linearization errors, the proposed PF method can accommodate widespread operating conditions due to fast-varying DER generation with guaranteed computational accuracy. Moreover, the proposed method is robust to bad data possibly existing in the measurement dataset to mitigate its adverse effects.



The linear solution, denoted by $\mathbf v^{tay}$, which is analytical, is attained by expressing \eqref{linearPF} in rectangular coordinates as:
\begin{equation} \label{InitialS}
\begin{bmatrix}
\mathbf A_r+\mathbf B_r & \mathbf A_x-\mathbf B_x\\\mathbf A_x+\mathbf B_x & -\mathbf A_r+\mathbf B_r
\end{bmatrix}
\begin{bmatrix}
\mathbf v_r^{tay} \\\mathbf v_x^{tay}
\end{bmatrix} = \begin{bmatrix}
\mathbf d_r \\ \mathbf d_x\end{bmatrix}
\end{equation}
where the subscripts $r$ and $x$ denote the real and imaginary parts of complex numbers, respectively. Then, we have
\begin{equation} \label{InitialS1}
\begin{bmatrix}
\mathbf v_r^{tay}  \\ \mathbf v_x^{tay} 
\end{bmatrix}=
\begin{bmatrix}\mathbf A_r+\mathbf B_r & \mathbf A_x-\mathbf B_x\\\mathbf A_x+\mathbf B_x & -\mathbf A_r+\mathbf B_r
\end{bmatrix}^{-1}
\begin{bmatrix}
\mathbf d_r \\ \mathbf d_x\end{bmatrix} 
\end{equation}

In the case of increases in accuracy loss from the Taylor-expansion-based approximation, a natural idea is to estimate the linearization errors, thereby maintaining the accuracy of the PF calculation for a wide range of operating conditions. Considering the linearization errors, the accurate solution for the AC nonlinear PF can be expressed by:
\begin{equation} \label{eq2_19}
\begin{bmatrix}
\mathbf V_{Nr} \\\mathbf V_{Nx}
\end{bmatrix}=
\begin{bmatrix}
\mathbf v_r^{tay} \\ \mathbf v_x^{tay}\end{bmatrix} +
\begin{bmatrix}
 \boldsymbol{\omega}  \\ \boldsymbol{\mu} 
\end{bmatrix}\quad\text{for brevity}\quad \mathbf x=\mathbf x^{tay} + 
 \mathbf e
\end{equation}
where $\mathbf v_{r}^{tay}$ and $\mathbf v_{x}^{tay}$ denote the real and imaginary Taylor-approximated voltage vector, respectively; $\mathbf e=[ \boldsymbol\omega,\boldsymbol\mu ]^\top$
denotes the linearized error vector introduced by the Taylor expansion-based approximation, with $\boldsymbol{\omega}$ and $\boldsymbol{\mu}$ as the real and imaginary error vectors, respectively.


The linearization error $\mathbf e$  is estimated by a regression model:
\begin{equation} \label{eq2_14}
\hat{\mathbf{e}}=
\mathbf w \odot \mathbf y + \mathbf b
\end{equation}
where $\mathbf{y}=\left[\Re (\mathbf V_0); \Im (\mathbf V_0);\mathbf P;\mathbf{Q}; \right]$ is the model input; $\mathbf w \in  \mathbb{R}^{2\mathcal{M}}$ denotes the regression coefficients, and $\mathbf b \in \mathbb{R}^{2\mathcal{M}}$ denotes the offset vector. 

Based on the historical dataset $\{\mathbf{x}_j,\mathbf{y}_j\}_{j=1}^M$, the errors are calculated from both the historical results and the corresponding linear solution by \eqref{InitialS}, and $ \mathbf e_j =\mathbf x_j -\mathbf x^{tay}_j$. Here $M$ denotes the number of training input-output pairs. Treat $\{\mathbf{e}_j,\mathbf{y}_j\}_{j=1}^M$ as the new dataset of the proposed regression model. Then, the objective of linear regression is formulated to minimize the mean square error over a set of training instances by determining $\mathbf w$ and $\mathbf b$:
\begin{equation}\label{obj_learning_zip}
\{\mathbf w, \mathbf b\}= \text{arg} \text{ min} \quad 
\sum_{j=1}^M\|{\mathbf{e}_j- \hat{\mathbf{e}}_j}\|^2_2,
\end{equation}
where $\hat{\mathbf{e}}_j$ denotes the estimation of the error vector for the $j$th data sample.

While traditional linear regression approaches in \eqref{obj_learning_zip} hinge on data collinearity and possible outliers existing in the dataset, a robust SVR-based error estimation method is further proposed for accuracy and robustness enhancement.
SVR is a machine learning technique that seeks a hyperplane that fits the data by minimizing the margin of errors. The strength of SVR lies in its ability to mitigate the influence of bad data by exclusively penalizing observations that fall outside of its defined tolerance range.
SVR is formulated as a convex optimization problem with the following objective function and constraints to determine $\mathbf w$ and $\mathbf b$\cite{awad2015SVR}:
\begin{align}\label{svr}
{\textnormal{min}}\quad
& \frac{1}{2}\Vert\mathbf w\Vert^{2}+C\sum\limits_{j=1}^{M}(\xi_{j}+\xi_{j}^{*}) \\
\textnormal{s.t.}\quad
&\mathbf e_{j} - \mathbf w\odot \mathbf y_{j}- \mathbf b\leq\epsilon+\xi_{j} \nonumber \\
\label{svr2}
&\mathbf{w}\odot\mathbf y_{j}  + \mathbf b - \mathbf e_{j}\leq\epsilon+\xi_{j}^{*} \\
& \xi, \xi^{*}\geq 0 \nonumber
\end{align}
where $C$ denotes the regularization coefficient for penalization, and $\epsilon$ denotes the pre-defined maximum deviation that the model is permitted to tolerate; slack variables  $\xi_{j}$ and  $\xi_{j}^{*}$ are introduced to address infeasible constraints in case of the existence of outliers.
\begin{figure}[!t]
   \centering
   \includegraphics[width=0.44\textwidth]{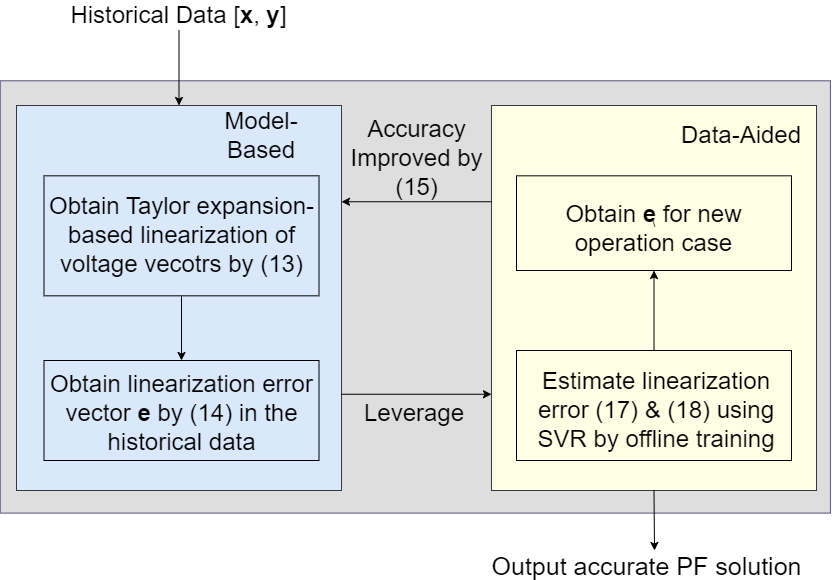}
 \caption{Flowchart of the proposed hybrid algorithm}
 \label{fig:2}
 \end{figure}
 
The procedure of the proposed hybrid PF algorithm, consisting of offline training and online execution, is illustrated in Fig.\ref{fig:2}. 
First, the Taylor expansion-based linear solution $\mathbf{x}^{tay}$ and approximation error vectors $\mathbf{e}$ are calculated from \eqref{InitialS} and \eqref{eq2_19}. 
By offline training, SVR is leveraged on the historical data to approximate the linearization error $\hat{\mathbf{e}}$ via the regression model
\eqref{eq2_14}. When executed online, the required time of the proposed hybrid method is minimal, allowing for real-time implementation. For new operating cases, the proposed method obtains accurate PF solution $\hat{\mathbf{x}}$ by:
\begin{equation}\label{final_form}
\hat{\mathbf{x}} = \mathbf{x}^{tay} + \hat{\mathbf{e}} 
\end{equation}
\vspace{-5pt}


\vspace{-8pt}
\section{Case Study}
\begin{figure}[!t]
\centering
\includegraphics[width=3.4in]{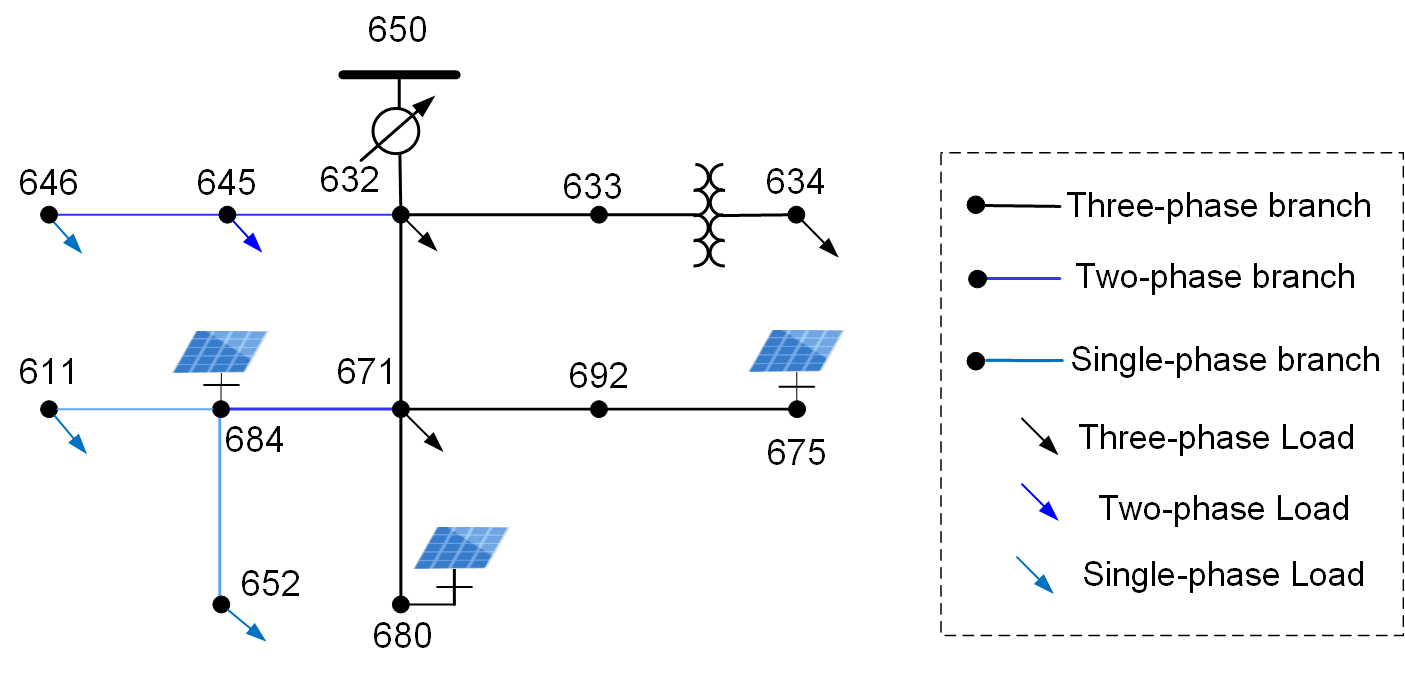}
\caption{The modified IEEE 13-node feeder with PV installation.}
\label{fig:3}
\vspace{-10pt}
\end{figure}


The proposed hybrid SVR-aided algorithm is tested on the unbalanced IEEE 13- and 123-bus distribution system \cite{feeder}. To consider scenarios with high DER penetration, these feeders are modified by adding photovoltaic (PV) solar panels. The modified 13-bus feeder with the PV installation is shown in Fig.\ref{fig:3}, while the details of the 123-bus test system can be found in \cite{YZ8753417}.
A variation of 20\% was assumed for all the DER generation, while 10\% is superposed to power load demands, which is used to generate the historical data in a wide range of operating conditions. The nonlinear PF results for unbalanced distribution systems are obtained by the iterative BFS method \cite{kersting2018distribution}. Moreover, random noises are introduced to the power injections as the input in the training datasets. 
The training and testing dataset contains 4000 and 1000 data samples, respectively. Gaussian noise with 10\% maximum errors for power injection is added to the training dataset as the measurement noises. 

Root mean square errors (RMSEs) of the estimated voltages relative to the true values are used to evaluate the computational accuracy. The RMSEs of voltage magnitudes and phase angles are calculated on each phase for all the test cases.

The effectiveness of the proposed method is gauged by comparing it with three approaches: model-based BFS, linear regression (LR), and SVR \cite{awad2015SVR}. 
All programs are implemented on a desktop computer with an Intel Core i9-10900K 3.70-GHz processor and 32 GB of RAM.

\begin{table}[!t]
\captionsetup{font=small}
\caption{\textsc{Comparison of RMSEs in Different Methods}}
\label{tab:table2}
\centering
\resizebox{\columnwidth}{!}{
\begin{tabular}{|c|c|c|c|c|c|c|}
\hline 
\multirow{2}{*}{\#Phase} & \multicolumn{3}{|c|}{Voltage Magnitude [p.u.]} & \multicolumn{3}{|c|}{Phase Angle [rad]}\\
\cline {2-7} &Ours& LR &SVR\cite{RobustThree9347375}& Ours& LR& SVR\cite{RobustThree9347375}\\
\hline A&   3.60e-5& 8.69e-5& 6.40e-3& 2.20e-3& 4.60e-3&1.18e-2\\
\hline B&   3.75e-4& 3.95e-4& 6.60e-3& 1.11e-4& 1.17e-4&5.94e-4\\
\hline C&   3.64e-4& 3.99e-4& 5.60e-3& 1.96e-4& 2.06e-4&1.20e-3\\
\hline
\end{tabular}
}
\end{table}


\begin{table}[!t]
\centering
\captionsetup{font=small}
\caption{\textsc{Average CPU Time For Test Cases}}\label{tab:table3}
\begin{tabular}{|c|c|c|c|c|}
\hline 
Proposed Method & w/o bad data [s] & w/ bad data [s]\\
\hline 
13-Bus Feeder & 0.0024& 0.0031\\
\hline
123-Bus Feeder & 0.112 & 0.126  \\
\hline
\end{tabular}
\end{table}

\subsection{Computation Accuracy and Time}

The computation accuracy of the proposed hybrid SVR-aided algorithm is compared with data-driven LR- and SVR-based approaches, and the scenario without bad data is considered here for preliminary analysis. 
Table \ref{tab:table2} lists the comparison in the RMSEs of voltages obtained by these methods in the 13-bus system. It can be seen that the RMSEs of voltage magnitudes of the proposed method are about 15 to 50 times lower than that of the SVR approach. Benefiting from the aid of the Taylor expansion, the proposed hybrid method achieves such significant accuracy enhancement over the pure data-driven LR- and SVR-based method.

The average computation time of the proposed method for all the test cases is listed in Table \ref{tab:table3}. Even with the interruption of bad data in the measurement dataset, the CPU time of the proposed robust method remains minimal, around 0.126 seconds for the 123-bus system. 
Thus, adopting the proposed algorithm for real-time implementation is promising, as the data-driven SVR model can be trained offline efficiently.


\subsection{Robustness Against Bad Data}

\begin{figure}[!t]
\centering
\includegraphics[width=3.4in]{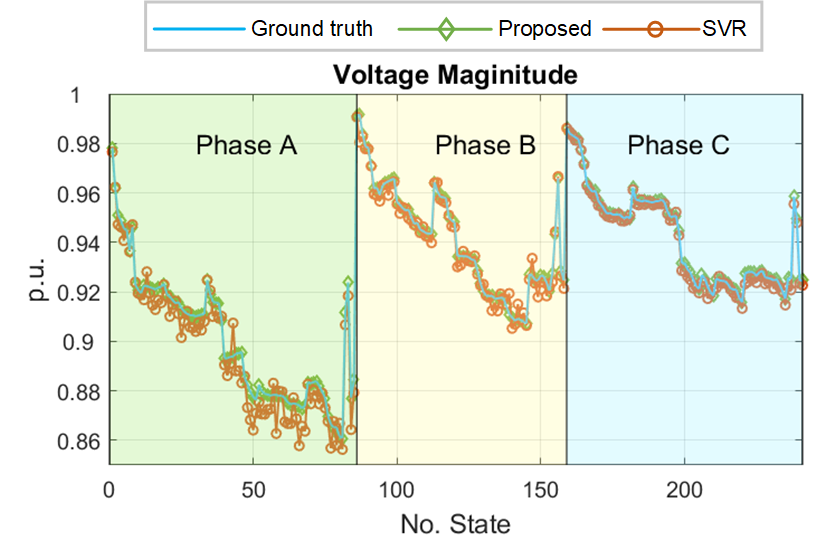}
\caption{Comparison of three-phase voltage magnitude for the 123-bus test feeder using SVR and the proposed methods.}
\label{fig:4}
\vspace{-10pt}
\end{figure}
\begin{figure}[!t]
\centering
\includegraphics[width=3.45in]{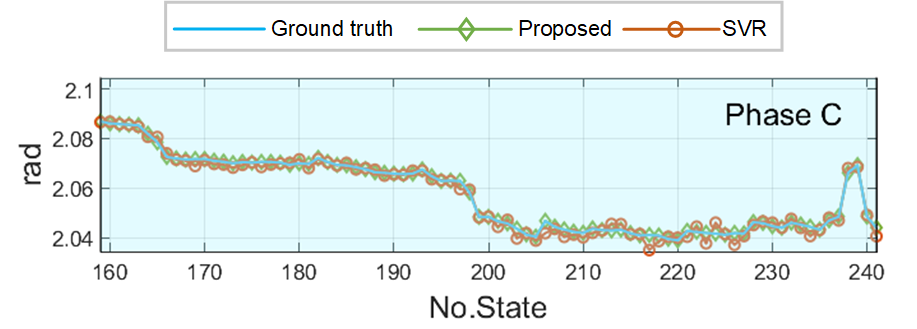}
\caption{Comparison of voltage angles using SVR and the proposed methods. Limited by space, only C-phase angles are illustrated.}
\label{fig:5}
\vspace{-5pt}
\end{figure}

The efficacy of the proposed SVR-aided algorithm is assessed in case bad data exists. 10\% of the training datasets are randomly selected to contain bad data \cite{Zhengshuo9626603}, and among them, totally $n_{bad}$ nodes are randomly selected as those with bad data. The following arrangements are adopted to test the robustness against bad data: for voltage magnitudes, the bad data is set as close to either 0 p.u. or larger than 3.0 p.u \cite{Zhengshuo9626603}; the power data is interrupted by injecting 1.5 times of measurement errors as the bad data. Moreover, set $n_{bad}=3$ for the 13-bus test feeder, while $n_{bad}=10$ for the 123-bus distribution feeder.

\begin{table}[!t]
\captionsetup{font=small}
\caption{\textsc{Robustness Comparison Against Bad Data}}
\label{tab:table5}
\centering
\resizebox{\columnwidth}{!}{
\begin{tabular}{|c|c|c|c|c|c|c|}
\hline 
\multirow{2}{*}{\#Phase} & \multicolumn{3}{c|}{Voltage Magnitude [p.u.]} & \multicolumn{3}{|c|}{Phase Angle [rad]}\\
\cline {2-7} &Ours& LR &SVR\cite{RobustThree9347375}& Ours& LR& SVR\cite{RobustThree9347375}\\
\hline A& 2.52e-4& 1.96e-2& 5.84e-3& 7.34e-4& 2.09e-4&8.13e-2\\
\hline B& 2.29e-5& 1.94e-2& 2.59e-3& 4.55e-6& 7.37e-6&1.02e-3\\
\hline C& 1.04e-5& 5.10e-3& 2.21e-3& 3.24e-5& 4.01e-6&8.12e-4\\
\hline
\end{tabular}
}
\vspace{-5pt}
\end{table}

\textbf{Compare with data-driven LR and SVR methods}. Fig.\ref{fig:4} and Fig.\ref{fig:5} plot the voltage magnitudes and phase angles on all the nodes attained by different methods in the 123-bus test feeder. The ground-truth values of voltages are also illustrated.
It is demonstrated that the proposed hybrid algorithm derives accurate voltage results with the existence of measurement outliers, while the SVR method is severely disturbed by the bad data.
The RMSE results of the proposed algorithm and data-driven methods are listed in Table \ref{tab:table5}. The proposed method exhibits robustness to bad data, as indicated by the lowest RMSE values. The RMSEs decrease from $5.84 \times 10^{-3}$ p.u. to $2.52 \times 10^{-4}$ p.u. for A-phase voltage. Thus, the proposed method realizes at least 20-fold accuracy enhancement compared to the pure data-driven SVR. This is because the participation of the Taylor-expansion formulation provides the learning interpretability for the subsequent regression task.

\textbf{Compare with model-based BFS}. For all the bad data-included data instances, the RMSEs of voltages are calculated for robustness evaluation. The proposed method is compared with the model-based BFS method in Table \ref{tab:table4} for the 123-bus system. The performance of the model-based BFS method is sensitive to bad data, and shown in Table \ref{tab:table4}, its accuracy significantly deteriorates. In contrast to BFS, the proposed method exhibits higher accuracy, with a remarkable 30-fold improvement in voltage magnitude and a 20-fold boost in voltage phase angles.

In summary, the proposed hybrid method attains much higher accuracy than other data-driven approaches in both scenarios with and without bad data. Moreover, differentiating from the BFS method without endowing robustness against bad data, the superior performance of the proposed method persists for measurements with about 30\% of bad data. 


\begin{table}[!t]
\captionsetup{font=small}
\caption{\textsc{Bad Data Impacts on BFS and Proposed Method}}
\label{tab:table4}
\centering
\begin{tabular}{|c|c|c|c|c|}
\hline 
\multirow{2}{*}{RMSEs} & \multicolumn{2}{c|}{Model-Based BFS} & \multicolumn{2}{c|}{Proposed Robust Method}\\
\cline {2-5} &Magn. [p.u.]&Angle [rad] & Magn.[p.u.] &Angle [rad]\\
\hline Phase A& 7.10e-3 & 1.41e-2&2.52e-4& 7.34e-4\\
\hline Phase B& 2.24e-2 & 1.27e-2& 2.29e-5& 4.55e-6\\
\hline Phase C& 1.32e-2 & 4.01e-3& 1.04e-5& 3.24e-5\\
\hline
\end{tabular}
\vspace{-10pt}
\end{table}

\section{Conclusion}
This paper proposes a novel Taylor-expansion-based SVR-aided robust PF algorithm in unbalanced distribution systems. The proposed hybrid model-based and data-driven algorithm is able to tackle unknown errors and bad data existing in the measurement dataset. This merit is vital for accommodating the random uncertain generation of DER units.
Comparative analysis with several model-based and data-driven methods shows that our algorithm achieves superior computational efficiency, with guaranteed accuracy and robustness against bad data. Our future work will investigate the proposed method's robustness against varying DER penetration levels and extend it to integrating ZIP load models and unknown topology.

\bibliographystyle{IEEEtran}
\bibliography{IEEEabrv,Citation}
\let\mybibitem\bibitem



\end{document}